\documentclass[twoside]{dis07}
\usepackage[latin1]{inputenc}
\usepackage[dvips]{graphicx,epsfig,color}
\usepackage{wrapfig,rotating}
\usepackage{amssymb,amsmath,array}

\newcommand{\be}{\begin{equation}}
\newcommand{\ee}{\end{equation}}
\newcommand{\bea}{\begin{eqnarray}}
\newcommand{\eea}{\end{eqnarray}}

\newcommand{\bfk}{\mbox{\boldmath $k$}}

\def\kt{k_\perp}
\newcommand{\bfp}{\mbox{\boldmath $p$}}
\def\bpo{{\bfp}_{\perp 1}}
\def\bpt{{\bfp}_{\perp 2}}

\newcommand{\bfq}{\mbox{\boldmath $q$}}
\newcommand{\bfP}{\mbox{\boldmath $P$}}

\def\ppo{p_{\perp 1}}
\def\ppt{p_{\perp 2}}
\def\pp{p_\perp}

\newcommand{\ua}{\uparrow}

\def\avk{\langle k_\perp ^2\rangle}
\def\avp{\langle p_\perp ^2\rangle}
\def\avk{\langle k_\perp ^2\rangle}
\def\avp{\langle p_\perp ^2\rangle}
\def\avPT{\langle P_T^2\rangle}

\def\T{_{_T}}
\def\C{_{_C}}

\begin{document}

\title{Transversity and Collins Functions: \\
from $e^+e^-\to h_1 h_2 X$ to  SIDIS Processes\footnote{Talk
  delivered by U.~D'Alesio at the ``15th International Workshop on 
  DIS'', DIS2007, April 16-21, 2007, Munich, Germany.}} 

\author{M.~Anselmino$^1$, M.~Boglione$^1$, U.~D'Alesio$^2$,
  A.~Kotzinian$^{1,3}$, F.~Murgia$^2$, A.~Prokudin$^1$ and C.~T\"urk$^1$ 
\vspace{.2cm}\\
1- Dipartimento di Fisica Teorica, Universit\`a di Torino and 
INFN, Sezione di Torino \\
Via P. Giuria 1, I-10125 Torino,  Italy
\vspace{.05cm}\\
2- Dipartimento di Fisica, Universit\`a di Cagliari and
          INFN, Sezione di Cagliari,\\
          C.P. 170, I-09042 Monserrato (CA), Italy
\vspace{.05cm}\\
3- Yerevan Physics Institute, 375036 Yerevan, Armenia, 
          JINR, 141980 Dubna, Russia \\
}

\maketitle

\vspace{-.35cm}
\begin{abstract}
We present~\cite{myurl} the first 
simultaneous extraction of the transversity distribution and 
the Collins fragmentation function,   
obtained
through a combined analysis of experimental data on azimuthal asymmetries in 
semi-inclusive deep inelastic scattering (SIDIS),
from the HERMES and COMPASS Collaborations, 
and in $e^+e^- \to h_1 h_2 X$ processes, from the Belle Collaboration.  
\end{abstract}

Among the three leading twist parton distributions, that contain basic
information on the internal structure of nucleons, 
the transversity
distribution is the most difficult to access. Due to its
chiral-odd nature it can only appear in
physical processes which require a quark helicity flip. 
At present the most accessible
channel is the SIDIS process with a 
polarized target, where the corresponding azimuthal asymmetry,
$A_{UT}^{\sin(\phi_S+\phi_h)}$,   
involves the transversity distribution coupled to
the Collins fragmentation function~\cite{Collins:1992kk}, also
unknown.   
Indeed it has received a lot of
attention in the ongoing experimental programs of HERMES
\cite{Airapetian:2004tw}, COMPASS \cite{Ageev:2006da}, and JLab
Collaborations. 

A crucial breakthrough in this strategy has recently been achieved 
with the independent measurement of the Collins function via the
azimuthal correlation observed in the two-pion production in $e^+e^-$
annihilation by the Belle Collaboration at KEK \cite{Abe:2005zx}. 

Let us start with the $e^+e^-\to h_1 h_2 X$ process. 
We choose the 
reference frame so that the elementary $e^+e^-\to q\bar q$  scattering
occurs in the $\hat{xz}$ plane, with the back-to-back quark and antiquark
moving along the $\hat z$-axis identified as the jet thrust axis. The
cross section corresponding to this process can be expressed 
as (see Ref.~\cite{Anselmino:2007fs}):  
\bea
\frac{d\sigma ^{e^+e^-\to h_1 h_2 X}}
{dz_1\,dz_2\,d^2\bpo\,d^2\bpt\,d\cos\theta}&=&
 \frac{3\pi\alpha^2}{2s} \, \sum _q e_q^2 \, \Big\{
 (1+\cos^2\theta)\,D_{h_1/q}(z_1,\ppo)\,D_{h_2/\bar q}(z_2,\ppt)
\Big. \nonumber \\ 
+ \,\Big. \frac{1}{4}
\,\sin^2\theta
& &\hspace*{-.8cm}
\Delta ^N D _{h_1/q^\ua}(z_1,\ppo)\,
 \Delta ^N D _{h_2/\bar q^\ua}(z_2,\ppt)\,\cos(\varphi_1 +
 \varphi_2)\Big\}\,, 
\label{belle}
\eea
where  $\varphi_i$ are the azimuthal 
angles identifying the direction of the 
observed hadron $h_i$ in the helicity frame of the fragmenting quark
$q$, $z_i$ and $\bfp_{\perp i}$ are the 
hadron light-cone momentum fractions and transverse
momenta, and $\theta$ is the scattering angle 
in the $e^+e^-\to q\bar q$ process.
$\Delta^N D_{h/q^\ua}(z,\pp)$ is the Collins function, 
also known as $H_1^\perp$ (see Ref.~\cite{Boer:1997mf}).  
To compare with data we have to $i)$ perform a change of angular variables
from $(\varphi_1,\varphi_2)$ to 
$(\varphi_1,\varphi_1+\varphi_2)$ and integrate 
over $p_{\perp 1}$, $p_{\perp 2}$, and over 
$\varphi_1$; $ii)$ normalize
the result to the azimuthal averaged cross section; $iii)$ take the
ratio $R$ of unlike-sign 
to like-sign pion-pair production:
 \be
 R \simeq 1+\cos(\varphi_1+\varphi_2)\,A_{12}(z_1,z_2)\>, \quad {\rm
 where}\quad 
  A_{12}(z_1,z_2)=\frac{1}{4}\,\frac{\langle \sin^2\theta \rangle}
 {\langle 1+\cos^2\theta \rangle}\,(P_U-P_L)\>,
 \label{Ra12}
 \ee
the angle $\theta$ is averaged over a range of values given by the
detector acceptance, 
 \be
 P_{U(L)}=\frac{\sum_q e^2_q \, \Delta ^N D_{\pi^+/q^\uparrow}(z_1)\,
 \Delta ^N D_{\pi^{-(+)}/\bar q^\uparrow}(z_2)}{\sum_q e^2_q D
 _{\pi^+/q}(z_1)\,  D _{\pi^{-(+)}/\bar q}(z_2)}\>,
 \label{pheno:pul}\quad {\rm and}
 \ee
 \be
 \Delta^ND_{h/q^\uparrow}(z) = \int d^2\bfp_\perp \Delta^N
 D_{h/q^\uparrow}(z,p_\perp) 
             = \int d^2\bfp_\perp \frac{2p_\perp}{z m_h} \;
 H_1^{\perp q}(z,p_\perp)  = 4 \; H_1^{\perp(1/2)q}(z)\>.
 \label{coll-mom}
 \ee
For fitting purposes, it is convenient to re-express $P_U$ and $P_L$ in terms
of favoured and unfavoured fragmentation functions (and similarly for
the $\Delta^ND$), 
\be
D_{\pi^+/u,\bar d} = D_{\pi^-/d,\bar u} \equiv D_{\rm fav} \,;\label{fav}
\;\;\; 
D_{\pi^+/d,\bar u} = D_{\pi^-/u,\bar d} = D_{\pi^\pm/s,\bar s} 
\equiv D_{\rm unf}\,. 
\ee
In addition, the Belle Collaboration presents the same set of data, analysed
in a different reference frame: following Ref.~\cite{Boer:1997mf}, one can
fix the $\hat z$-axis as given by the direction of the observed hadron $h_2$
and the $\hat{xz}$ plane as determined by the lepton and the $h_2$ directions.
An azimuthal dependence of the other hadron $h_1$ with respect to this
plane has been measured.  
In this configuration the corresponding ratio becomes 
\be
R \simeq 1+\cos(2\,\phi_1)A_0(z_1,z_2)\,,\;\; 
 \;\;
A_0(z_1,z_2)=\frac{1}{\pi} \, \frac{z_1\,z_2}{z_1^2+z_2^2}
\, \frac{\langle \sin^2\theta_2 \rangle}
{\langle 1+\cos^2\theta_2 \rangle}\,(P_U-P_L )
\label{A0}\,.
\ee

Let us now consider the SIDIS process $\ell \, p \to \ell \, h \, X$.
 We take, in the $\gamma ^* -p$ c.m. frame,  
the virtual photon and the
proton colliding along the $\hat z$-axis with momenta $\bfq$ and $\bfP$
respectively, and the leptonic plane to coincide with the $\hat{xz}$ plane.

To single out the spin dependent part of the fragmentation
of a transversely polarized quark we consider  
the $\sin(\phi_S +\phi_h)$ weighted asymmetry (at ${\cal O}(\kt/Q)$):
\bea
\label{sin-asym}
&&A^{\sin (\phi_S + \phi_h)}_{{UT}} = 
2 \, \frac{\int d\phi_S \, d\phi_h \,
[d\sigma^\uparrow - d\sigma^\downarrow] \, \sin(\phi_S +\phi_h)}
{\int d\phi_S \, d\phi_h \,
[d\sigma^\uparrow + d\sigma^\downarrow]}\\
&&= 
\frac{\displaystyle  \sum_q e_q^2  \! \! \int \! \!{d\phi_S d\phi_h d^2
\bfk _\perp}\Delta _T q (x,\kt) \!
\frac{d (\Delta {\hat \sigma})}{dy}
\Delta^N\! D_{h/q^\ua}(z,\pp) \sin(\phi_S + \varphi +\phi_q^h)
\sin(\phi_S +\phi_h) } {\displaystyle \sum_q e_q^2 \int {d\phi_S
d\phi_h \, d^2 \bfk _\perp} f_{q/p}(x,k _\perp)
\frac{d\hat\sigma}{dy} D_{h/q}(z,p_\perp) } \, \cdot\nonumber
\eea
In the above equation $\Delta_Tq(x,\kt)$ is the unintegrated transversity
distribution, 
$d{\hat \sigma}/dy$ is the planar unpolarized elementary cross section
and $\frac{d(\Delta {\hat \sigma})}{dy} =
\frac{4\pi\alpha^2}{sxy^2}\,(1-y)
$.
The $\sin(\phi_S + \varphi +\phi_q^h)$ azimuthal dependence in
Eq.~(\ref{sin-asym}) arises from the combination of the phase factors in the
transversity distribution function, in the non-planar $\ell\,q \to \ell\,q$
elementary scattering amplitudes, and in the Collins fragmentation
function (see Ref.~\cite{Anselmino:2007fs} and \cite{Anselmino:2005sh}).  
We assume 
\be
 f_{q/p}(x,\kt)=
   f_{q/p}(x)\;\frac{e^{-{\kt^2}/\avk}}{\pi\avk}\;,\hspace*{0.5cm}
D_{h/q}(z,\pp)=D_{h/q}(z)\;\frac{e^{-\pp^2/\avp}}{\pi\avp}\;,
\label{unpfrag}
\ee
where $f_{q/p}(x)$ and $D_{h/q}(z)$ are the usual integrated parton
distribution and fragmentation functions  
and the average values of $\kt^2$ and $\pp^2$ are taken from
Ref.~\cite{Anselmino:2005nn}:
$\langle \kt^2 \rangle = 0.25 \> \textrm{GeV}^2$, 
$\langle \pp^2 \rangle = 0.20 \> \textrm{GeV}^2$. 
For the transversity and the Collins functions  
we choose 
\be
\Delta_T q(x, \kt) =
\frac{1}{2} \, {\cal N}^{T}_q(x)\,
\left[f_{q/p}(x)+\Delta q(x) \right] \;
\frac{e^{-{\kt^2}/{\avk}}}{\pi \avk} \label{tr-funct} \,, 
\ee
\be 
\Delta^N D_{h/q^\uparrow}(z,\pp) = 2\,{\cal N}^{C}_q(z)\;
D_{h/q}(z,p_\perp)
\,\sqrt{2e}\,\frac{p_\perp}{M}\,e^{-{p_\perp^2}/{M^2}}\,,
\label{coll-funct}
\ee
\be
{\cal N}^{T}_q(x)= N^{T}_q \,x^{\alpha} (1-x)^{\beta} \,
\frac{(\alpha + \beta)^{(\alpha +\beta)}} {\alpha^{\alpha} \beta^{\beta}}\,,
\;\;{\cal N}^{C}_q(z)= N^{C}_q \, z^{\gamma} (1-z)^{\delta} \,
\frac{(\gamma + \delta)^{(\gamma +\delta)}}
{\gamma^{\gamma} \delta^{\delta}}\,, 
\ee
with $|N^{T}_q|, \> |N^{C}_q| \le 1$ and where $\Delta q$ is the helicity
distribution. 

Notice that our parameterizations are devised in such a way that the
transversity distribution function and the Collins function
automatically obey their proper bounds. 
 
By insertion of the above expressions into Eq.~(\ref{sin-asym}), we obtain
\be
A^{\sin (\phi_S+\phi_h)}_{{UT}} =
\frac{\displaystyle  \frac{P_T}{M}\frac{1-y}{s x y^2}
\sqrt{2e} \frac{\avp ^2 \C}{\avp}
\frac{e^{-P_T^2/\avPT \C}}{\avPT ^2 \C} \sum_q e_q^2 
 {\cal N}^{\T}_q(x)
\left[f_{q/p}(x)+\Delta q(x) \right]
{\cal N}^{\C}_q(z)
D_{h/q}(z)}
{ \displaystyle \frac{e^{-P_T^2/\avPT}}{\avPT} 
\frac{[1+(1-y)^2]}{s x y^2}\,
 \sum_q e_q^2 \, f_{q/p}(x)\; D_{h/q}(z)}\;,
\label{sin-asym-final}
\ee
\be
{\rm with}\;\,
\avp \C= \frac{M^2 \avp}{M^2 +\avp}\,, \quad\quad
 \avPT=\avp+z^2\avk \,,\quad\quad
\avPT \C=\avp \C+z^2\avk\,.
\ee

Using the above expressions for $\Delta_Tq$ and
$\Delta^ND_{\pi/q^\uparrow}$ both in
$A^{\sin(\phi_S+\phi_h)}_{{UT}}$, Eq.~(\ref{sin-asym-final}), and
in $A_{12}$, Eq.~(\ref{Ra12}), we can fix all free parameters by  
performing a best fit of the HERMES, COMPASS and Belle data. We 
checked that using $A_0$ instead of $A_{12}$ leads to a consistent
extraction (see Ref.~\cite{Anselmino:2007fs} for details).

\begin{figure}[ht]
\vspace*{-0.2cm}
\begin{center}
\hspace*{1cm}
\includegraphics[width=0.3\textwidth,height=0.28\textwidth,bb= 10 140 540 660,
angle=-90]
{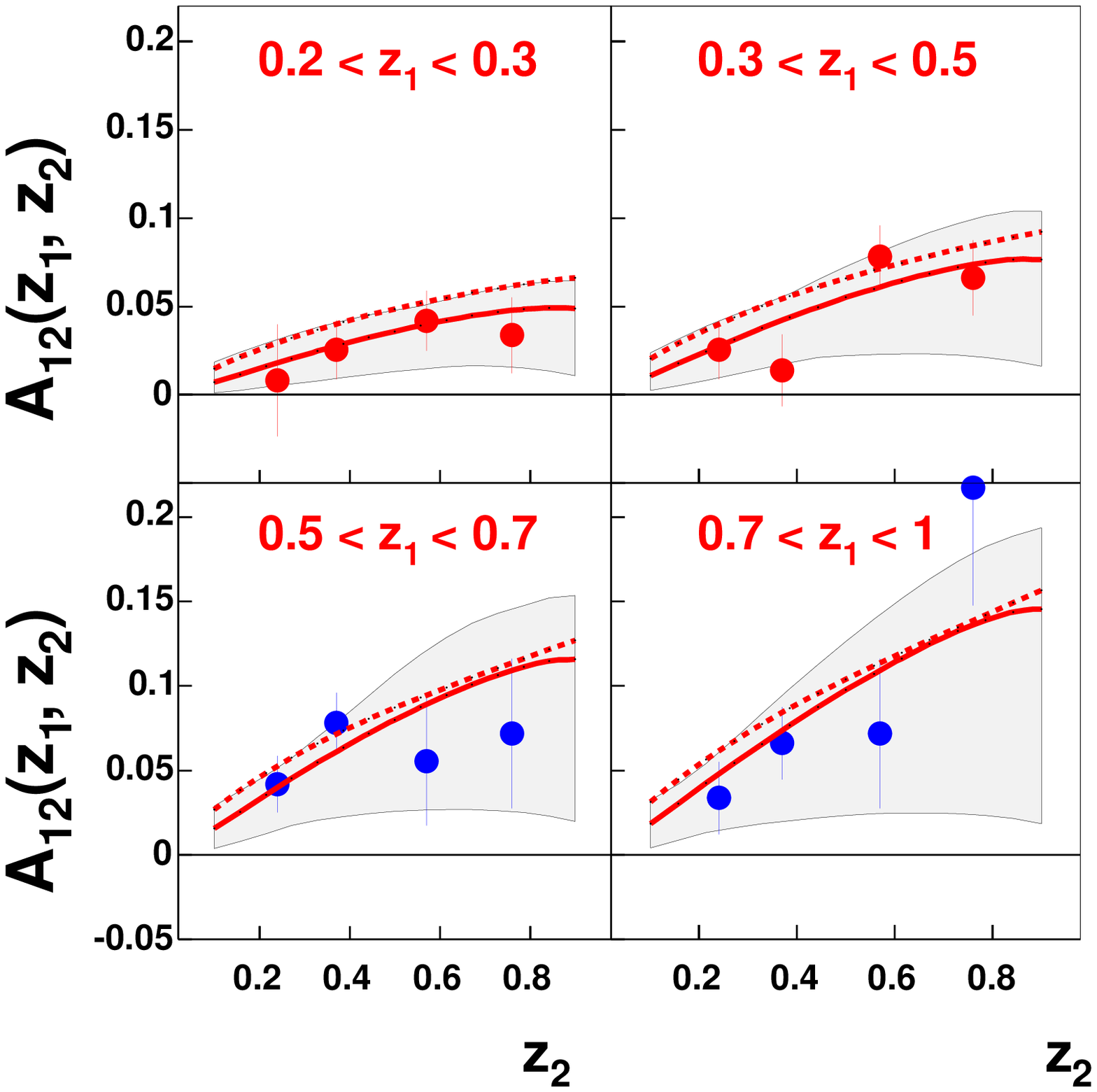} 
\includegraphics[width=0.3\textwidth,height=0.28\textwidth,bb= 10 140 540 660,
angle=-90]
{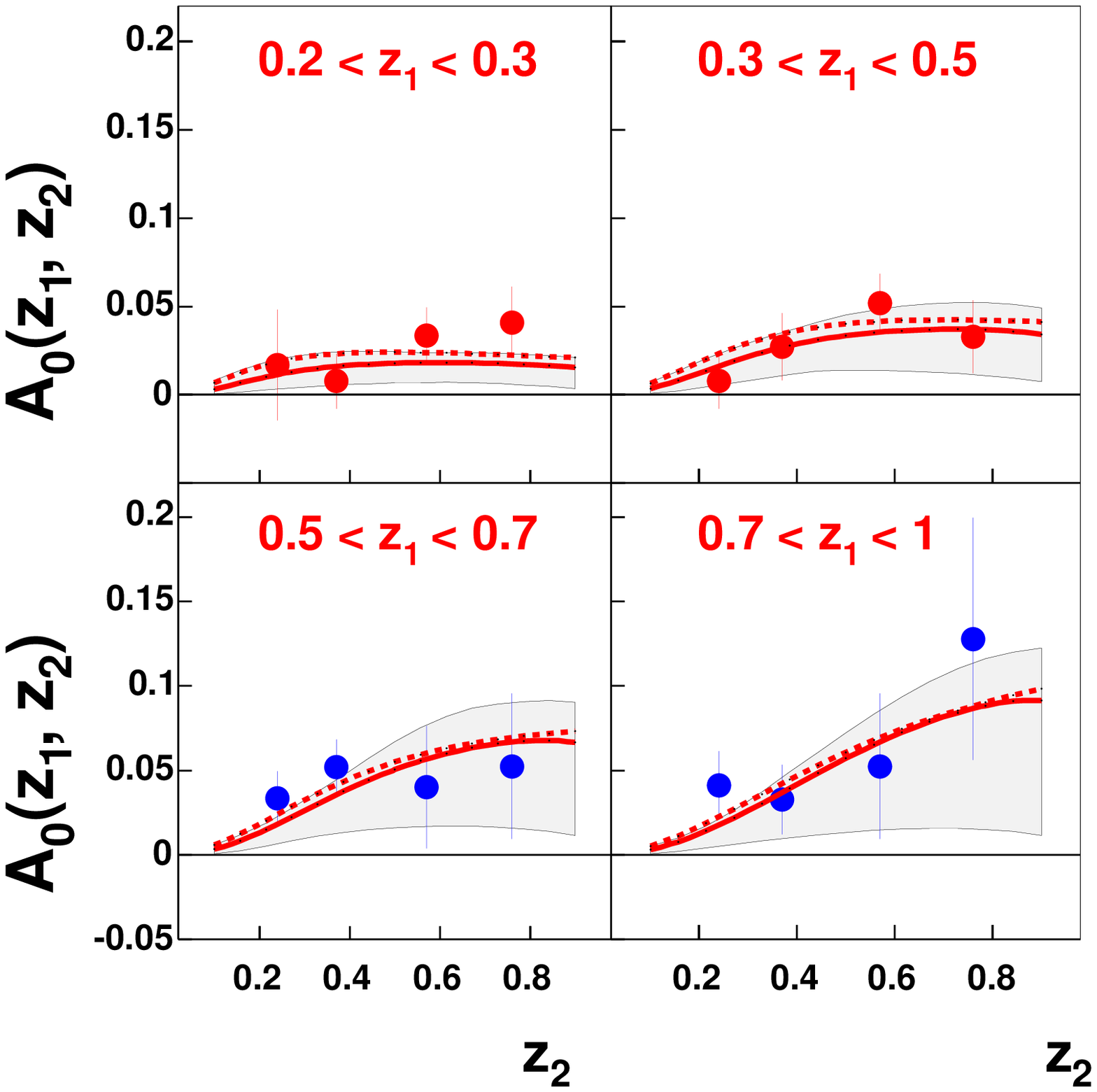}
\end{center}
\caption{\label{fig:belle}
Data on two different azimuthal correlations in unpolarized
$e^+e^- \to h_1 h_2 \, X$ processes, as measured by Belle Collaboration
\cite{Abe:2005zx}, compared to the curves obtained from our fit. 
The solid (dashed) lines correspond to the global fit obtained including  
the $A_{12}$($A_0$) asymmetry; 
the shaded area corresponds to the theoretical uncertainty on the 
parameters.
}
\end{figure}

\begin{figure}[h!t]
\vspace*{-.3cm}
\begin{center}
\includegraphics[width=0.28\textwidth,bb= 10 140 540 660,angle=-90]
{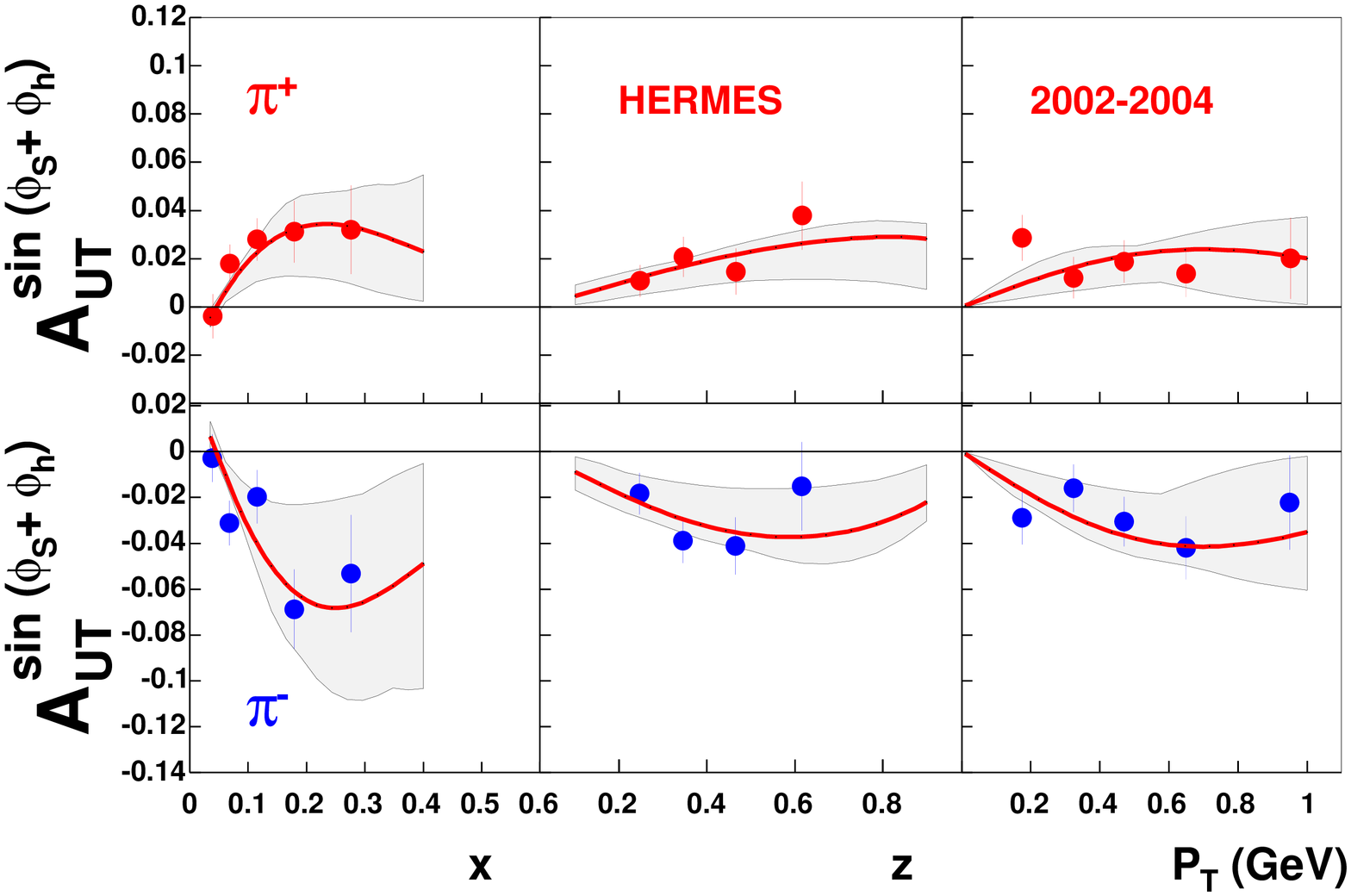} \hskip 2cm
\includegraphics[width=0.28\textwidth,bb= 10 140 540 660,angle=-90]
{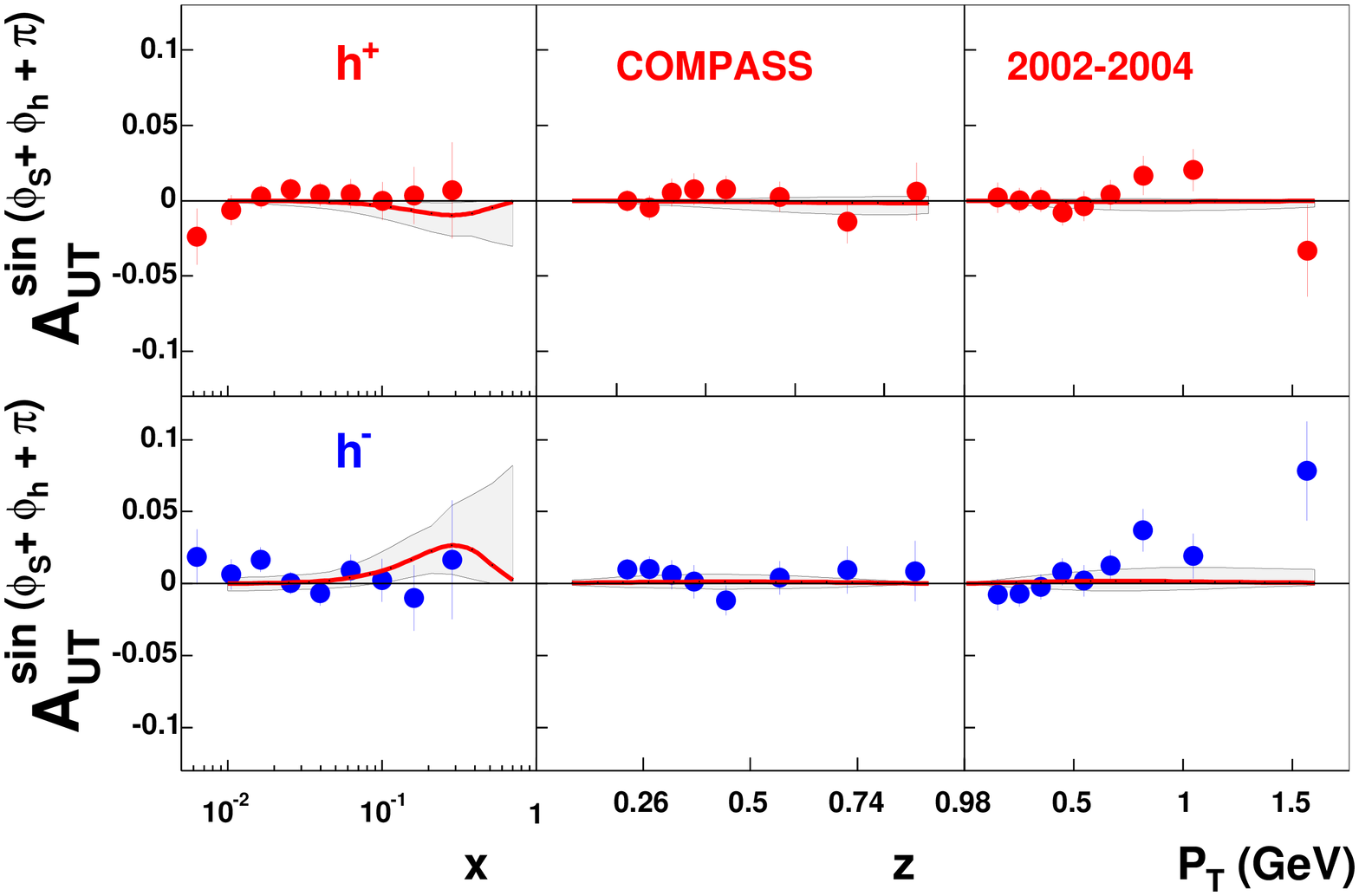} 
\end{center}
\caption{\label{fig:hermes}
Our results compared with 
HERMES data \cite{Airapetian:2004tw} on  
$A_{{UT}}^{\sin(\phi_S+\phi_h)}$ for $\pi^\pm$ production (left panel)
and COMPASS data on $A_{{UT}}^{\sin(\phi_S+\phi_h)}$, 
for the production of
positively and negatively charged hadrons off
a deuterium target~\cite{Ageev:2006da} (right panel). 
}
\end{figure}

\begin{figure}[ht]
\vspace*{-.0cm}
\begin{center}
\includegraphics[width=0.28\textwidth,bb= 10 140 540 660,angle=-90]
{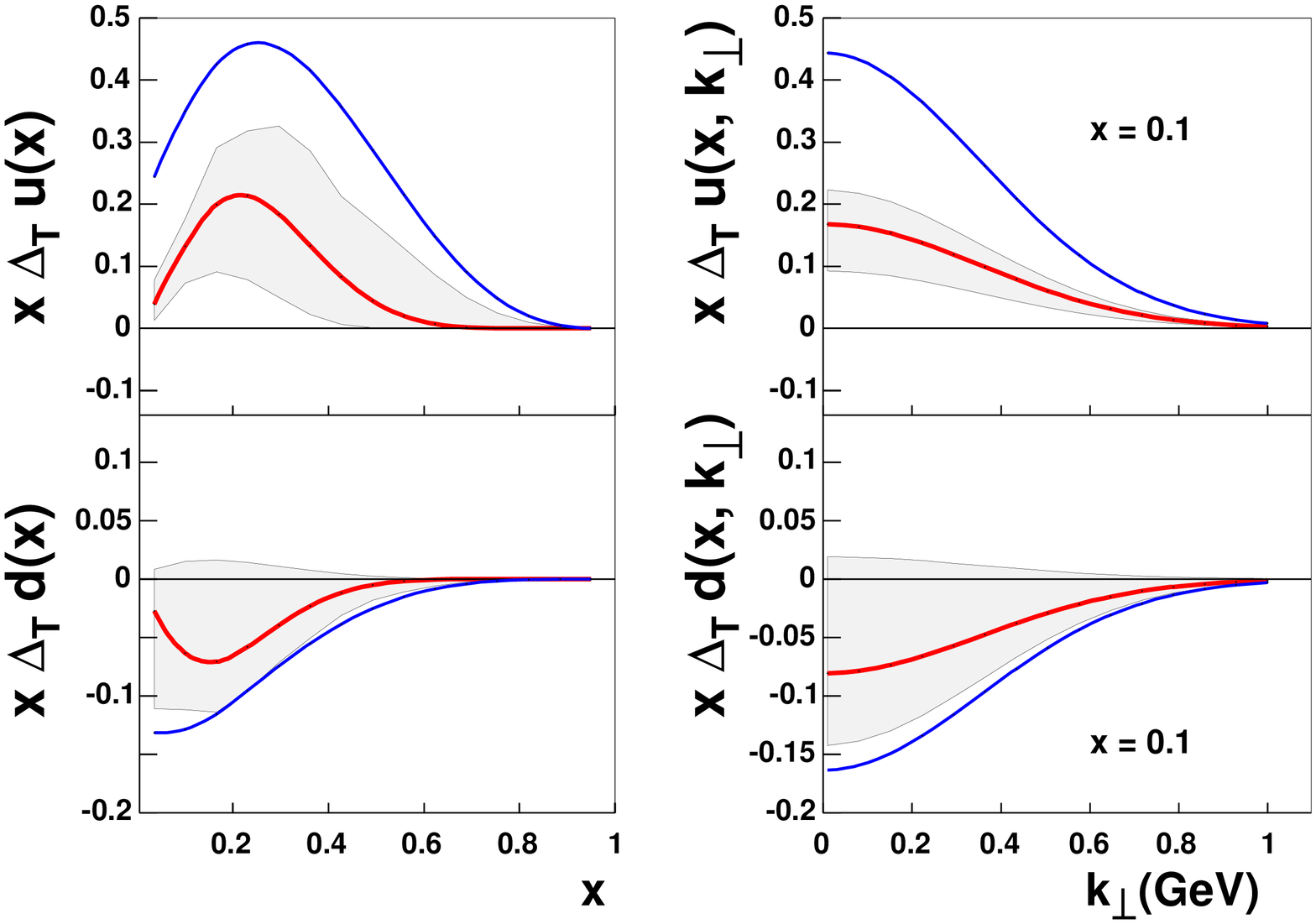}\hspace*{1.8cm}
\includegraphics[width=0.28\textwidth,,bb= 10 140 540 660,angle=-90]
{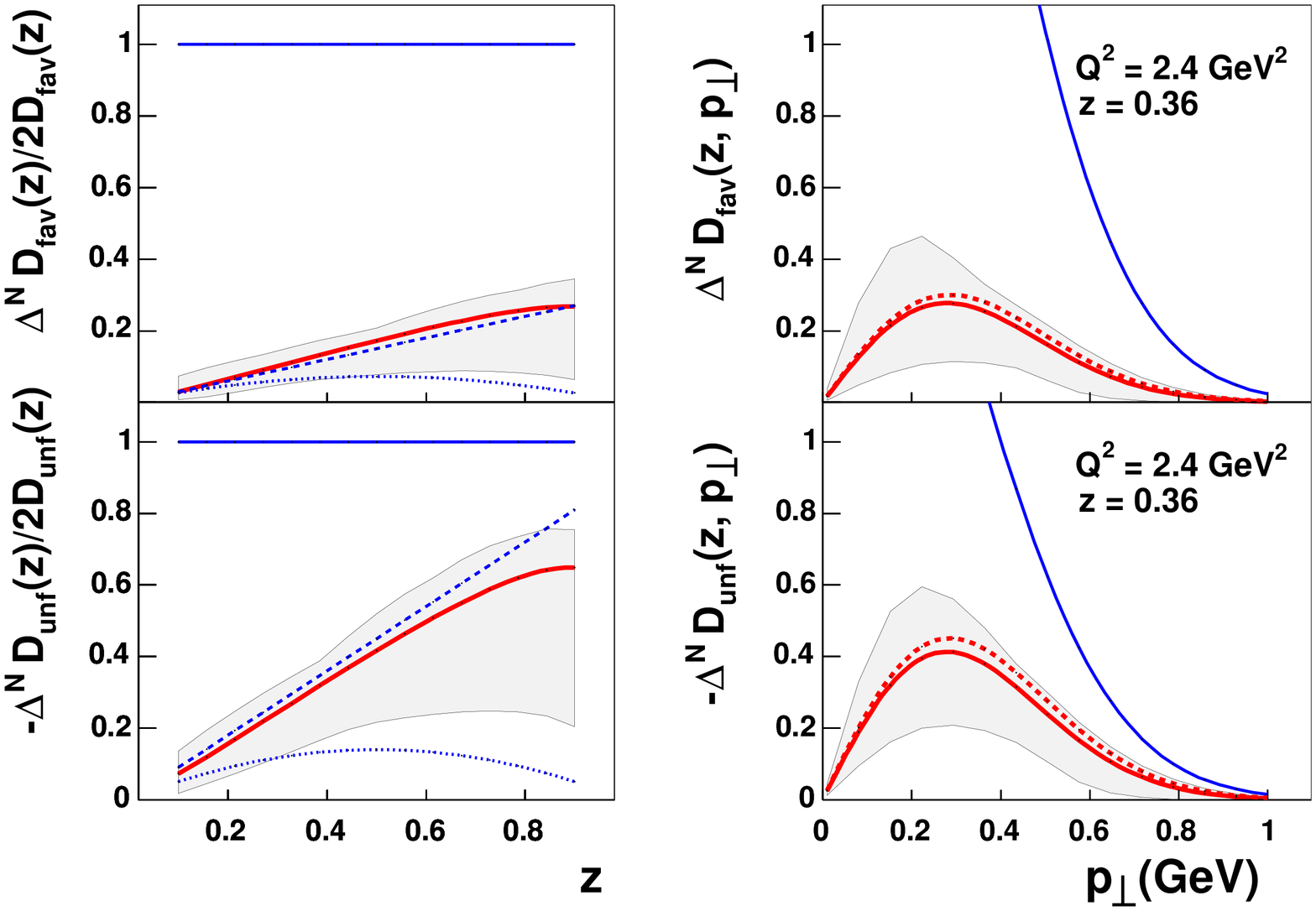}
\end{center}
\caption{\label{fig:transv}
First panel:
$x\,\Delta _T u(x)$ (upper plot) and $x\,\Delta _T d(x)$ (lower plot), 
vs. $x$ at $Q^2 = 2.4$ GeV$^2$. The Soffer bound 
is also shown for comparison (bold blue line).
Second panel: $x\,\Delta _T u(x,\kt)$ (upper plot) and $x\,\Delta _T d(x,\kt)$
(lower plot), vs. $\kt$ at a fixed value of $x$. 
Third panel: the $z$ dependence of the moment of the 
Collins functions, Eq.~(\ref{coll-mom}),  
normalized to twice the unpolarized fragmentation functions; 
also shown the results of Refs.~\cite{Efremov:2006qm} (dashed line) 
and \cite{Vogelsang:2005cs} (dotted line). Fourth panel:  the
$\pp$ dependence of the Collins functions.
}
\end{figure}

Our results are collected in Figs.~\ref{fig:belle}, \ref{fig:hermes}
where we present a comparison of our 
curves with the data. Figure~\ref{fig:transv} shows our extracted
transversity distributions and Collins functions. 

Summarizing, our global analysis of present data
on azimuthal asymmetries measured in SIDIS and $e^+e^-\to
\pi\pi X$ allows to get quantitative estimates of both the transversity and
the Collins function. In particular, we find:
$i)$ $|\Delta_Tu|>|\Delta_Td|$, and  both smaller than the
corresponding Soffer bound; $ii)$ $\Delta_T u$
tightly constrained by HERMES data alone, whereas COMPASS data 
help in constraining the transversity for $d$ quarks; 
$iii)$ unfavoured Collins functions larger in size (and opposite in
sign) than the favoured ones. 

\begin{footnotesize}

\end{footnotesize}


\end{document}